\documentclass[preprintnumbers, prd, twocolumn, showpacs,floatfix,preprintnumbers,
superscriptaddress, nofootinbib]{revtex4}
\usepackage{graphicx}
\usepackage{epsfig}
\usepackage{bm}
\usepackage{amssymb}
\usepackage{float}
\usepackage{amsmath}
\usepackage{subfigure}
\usepackage{dcolumn}
\usepackage{cancel}
\usepackage[colorlinks]{hyperref}
\usepackage[usenames,dvipsnames]{color}
\hypersetup{
     breaklinks=true,
    pdfstartview={FitH},    % fits the width of the page to the window
    colorlinks=true,       % false: boxed links; true: colored links
    linkcolor=blue,          % color of internal links
    citecolor=red,        % color of links to bibliography
    filecolor=magenta,      % color of file links
    urlcolor=blue,           % color of external links
    anchorcolor=green,      % Color for anchor text
    linktocpage=true
}

\def\doi{http://doi.org}

\begin{document}

\title{A new parametrization for dark energy density and future deceleration}

\author{Abdulla Al Mamon}
\email{abdulla.physics@gmail.com}
\affiliation{Department of Mathematics, Jadavpur University, Kolkata-700032, West Bengal, India}

\newcommand{\be}{\begin{equation}}
\newcommand{\ee}{\end{equation}}
\newcommand{\bea}{\begin{eqnarray}}
\newcommand{\eea}{\end{eqnarray}}
\newcommand{\bc}{\begin{center}}
\newcommand{\ec}{\end{center}}
%%%%%%%%%%%%%%%%%%%%%%%%%%%%%%%%%%%%%%%%%%%%%%%%%%%%%%%%%%%%%%%%%%%%%%%%%%
%%%%%%%%%%%%%%%%%%%%%%%%%%%%%%%%%%%%%%%%%%%%%%%%%%%%%%%%%%%%%%%%%%%%%%%%%%
\begin{abstract}
In this work, we have proposed a general dark energy density parametrization to study the evolution of the universe. We have also constrained the model parameters using the combination of Type Ia supernova (SNIa), baryonic acoustic oscillations (BAO), cosmic microwave background radiation (CMB) and observational $H(z)$ datasets. For the $H(z)$ dataset, we have used the direct observations of the Hubble rate, from the radial BAO size and the  cosmic chronometer methods. Our result indicates that the SNIa+$H(z)$+BAO/CMB dataset does not favour the $\Lambda$CDM model at more than $2\sigma$ confidence level. Furthermore, we have also measured the percentage deviation in the evolution of the normalized Hubble parameter for the present model compared to a $\Lambda$CDM model, and the corresponding deviation is found to be $4-5\%$ at low redshifts ($z\sim 0.5$). Finally, we have also investigated whether the deceleration parameter $q$ may have more than one transition during the evolution of the universe. The present model shows a transient accelerating phase, in which the universe was decelerated in the past and is presently accelerating, but will return to a decelerating phase in the near future. This result is in great contrast to the $\Lambda$CDM scenario, which predicts that the cosmic acceleration must remain forever.
\end{abstract} 
%%%%%%%%%%%%%%%%%%%%%%%%%%%%%%%%%%%%%%%%%%%%%%%%%%%%%%%%%%%%%  
\maketitle 
\section{Introduction}
%%%%%%%%%%%%%%%%%%%%%%%%%%%%%%%%%%%%%%%%%%%%%%%%%%%%%%%%%%
Various independent cosmic observations \cite{SN,Riess:1998cb,LSS,Seljak:2004xh,Planck:2015xua,planckr, Ade:2014xna,  Komatsu:2010fb, Hinshaw:2012aka,dvbc} have strongly confirmed that the present universe experiences an accelerated expansion. The exotic matter content responsible for such a certain phase of evolution of the universe is popularly referred to as ``dark energy". Many  dark energy models were proposed in the literatures, for a recent review, one can look into Refs. \cite{3,4,4a,bambaderev}. In the context of dark energy, the Einstein cosmological constant $\Lambda$ is the simplest way to explain the observed expansion measurements. The so-called concordance $\Lambda$CDM ($w_{\Lambda}=-1$) model is the model that best agrees with cosmological data \cite{Planck:2015xua}. Despite a very good agreement with data, the $\Lambda$CDM model can not escape from the cosmological {\it coincidence} and the {\it fine tuning} problems \cite{sw, Steinhardt} and is still a challenging problem in cosmology.\\
%%%%%%%%%%%%%%%%%%%%%%%%%%%%%%%%%%%%%%%%%%%%%%%%%%%%%%%%%%%%%%%%%%%%%%%
\par Going beyond the cosmological constant where the dark energy density is constant throughout
the evolution of the universe, there are several approaches to model the dark energy evolution \cite{4}. One best way is to construct parametrizations of the dark energy equation of state parameter \cite{cpl1,cpl2,lin1,lin2} or the dark energy density \cite{dedywang,dedmaor,dedwang,dedaam} as a function of scale factor or redshift, and then confront such parametrizations to the cosmological data. However, such models are more consistent with the present observational constraints for some restrictions on model parameters and search is still on for finding a suitable cosmologically viable model of dark energy. Recently, Zhao et al. \cite{zhao} reported that the dynamical dark energy is preferred over the cosmological constant model from recent observations at the $3.5\sigma$ confidence level, although the Bayesian evidence for the dynamical dark energy is insufficient to favour it over constant dark energy. This clearly motivates theoreticians to put further constraint on dark energy behaviour. In the present work, we have proposed a spatially flat FRW universe where the dark energy and the cold dark matter evolve independently. Specifically, we have considered a general dark energy density parametrization which varies with the cosmic evolution. The nature of this parametrization is characterized by dimensionless real parameters $\alpha$ and $n$. For different choices of $\alpha$ and $n$, one can recover other popular dark energy density parametrizations (see section \ref{sec2}). In this paper, we have used the recent cosmic chronometers dataset along with the estimation of the local Hubble parameter value as well as the standard dark energy probes, namely the SNIa, BAO and CMB measurements to study the different properties of this model extensively. Under this scenario, we also made an attempt to explain not only the present accelerated expansion phase but also the past decelerated phase of the universe and further made a prediction about the future evolution of the universe. Our analysis shows the evolution of the universe from an early decelerated to the late-time accelerated phase and it also predicts future decelerating phase. In addition, the present study also indicates that the $\Lambda$CDM model is not compatible at $2\sigma$ confidence level for the SNIa+$H(z)$+BAO/CMB dataset. \\
%%%%%%%%%%%%%%%%%%%%%%%%%%%%%%%%%%%%%%%%%%%%%%%%%%%%%%%%%%%%%%%%%%%%%%%%%%%
%%%%%%%%%%%%%%%%%%%%%%%%%%%%%%%%%%%%%%%%%%%%%%%%%%%%%%%%%%%%%%%%%%%%%%%%%%%%%%%%%%%%%%%%%
\par The paper is organized as follows. In the next section, we have discussed the present cosmological model. In Section \ref{data}, we have described the observational dataset and analysis methodology, while in section \ref{result} we have presented the results of this analysis. Finally, the summary of the work is presented in section \ref{conclusion}.
%%%%%%%%%%%%%%%%%%%%%%%%%%%%%%%%%%%%%%%%%%%%%%%%%%%%%%%%%%%%%%%%%%%%
%%%%%%%%%%%%%%%%%%%%%%%%%%%%%%%%%%%%%%%%%%%%%%%%%%%%%%%%%%%%%%%%
\section{Cosmological Model}\label{sec2}
%%%%%%%%%%%%%%%%%%%%%%%%%%%%%%%%%%%%%%%%%%%%%%%%%%%%%%%%%%%%%%%%%%%%%%%%%%%%%%%%%  
In this section, we have provided the basic equations of a general cosmological scenario. Throughout the work, we have considered the spatially flat FRW space-time of the form 
\begin{equation}
ds^{2} = dt^{2} - a^{2}(t)[dr^{2} + r^{2}(d{\theta}^{2} + sin^{2}\theta d{\phi}^{2})]
\label{eq:2.2}
\end{equation}
where, $a(t)$ is the scale factor of the universe, which is set to $1$ at the present epoch for simplicity and $t$ is the cosmic time. For a spatially flat FRW universe, the Einstein field equations can be written as,
\be\label{fe1}
3H^{2}=\rho_{DM}+\rho_{DE}
\ee
\be\label{fe2}
2{\dot{H}} + 3H^{2}=- p_{DE}
\ee
where $H=\frac{\dot{a}}{a}$ is the Hubble function and an dot implies differentiation with respect to the cosmic time $t$. In the above equation, $\rho_{DM}$ represents the energy density of the dust matter while $\rho_{DE}$ and $p_{DE}$ represent the energy density and pressure of the dark energy component respectively. It is noteworthy that we have chosen the natural units ($8\pi G = c = 1$) throughout this paper. \\
%%%%%%%%%%%%%%%%%%%%%%%%%%%%%%
\par One can now write the conservation equation of the dark energy sector and the one of the matter sector as
\be 
{\dot{\rho}}_{DE}+3H(p_{DE}+\rho_{DE})=0
\ee
\be
{\dot{\rho}}_{DM}+3H\rho_{DM}=0 
\ee
Solving the above equation, we have found the evolution of $\rho_{DM}$ as
\be\label{eqrdmz}
\rho_{DM}(z)=\rho_{DM0}(1+z)^3 
\ee
where $\rho_{DM0}$ denotes the present  matter energy density and $z=\frac{1}{a}-1$ is the redshift parameter.\\
%%%%%%%%%%%%%%%%%%%%%%%%%%%%%%%%%%%%%%%%%%%%%%%%%%%%%%%%%%%%%%%
%%%%%%%%%%%%%%%%%%%%%%%%%%%%%%%%%%%%%%%%%%%%%%%%%%%%%%%%%%%%%%%
\par In the present work, we have proposed a simple dark energy density parametrization that exhibits dynamical behaviour with the evolution of the universe. Our primary goal is to investigate this model with current cosmological data. To examine the nature of dark energy, we have proposed the following functional form for the evolution of $\rho_{DE}(z)$ given by
\be\label{eans1}
\rho_{DE}(z)=\rho_{DE0}\Big[1+\alpha {\Big(\frac{z}{1+z}\Big)^n} \Big],~~~n=2
\ee
where $\rho_{DE0}$ and $\alpha$ denote the present dark energy density and free parameter of the model respectively. One important advantage of this choice (as given in equation (\ref{eans1})) is that it reduces to the flat $\Lambda$CDM ($\rho_{DE}=\rho_{DE0}=$ constant) model for $\alpha=0$. So, the free parameter $\alpha$ is a good indicator of deviation of the present dark energy model from the $\Lambda$CDM model. Note that the above form of $\rho_{DE}(z)$ is similar to the parametrization of $\rho_{DE}(z)=\rho_{DE0}\Big[1+\alpha {\Big(\frac{z}{1+z}\Big)} \Big]$ \cite{dedwang}, if we put $n=1$ in equation (\ref{eans1}). Hence, the new parametrization of $\rho_{DE}(z)$ reduces to other cosmological models for some specific choices of $n$ and $\alpha$, and also shows a bounded behaviour in the redshift range, $-1<z<\infty$. Of course, $n$ may assume any real value, but we have found that for $n=2$, the expression of the dark energy density proposed here gives interesting consequences, as discussed in section \ref{result}. So, in the present work, we have confined our investigation to $n=2$ only.\\ 
%%%%%%%%%%%%%%%%%%%%%%%%%%%%%%%%%%%%%%%%%%%%%%%%%%%%%%%%%%%%%%%%%%%
\par With the help of equations (\ref{fe1}), (\ref{eqrdmz}) and (\ref{eans1}), the expression for the Hubble parameter for this model is obtained as 
\be\label{eqHz}
H(z)=H_{0}{\Big[\Omega_{DM0}(1+z)^3 + \Omega_{DE0}\Big(1+\alpha {\Big(\frac{z}{1+z}\Big)^2}\Big) \Big]}^{\frac{1}{2}}
\ee
which is equivalent to the $\Lambda$CDM model for $\alpha=0$. In the above equation, $H_{0}$, $\Omega_{DM0}=\frac{\rho_{DM0}}{3H^2_0}$, and $\Omega_{DE0}=\frac{\rho_{DE0}}{3H^2_0}$ denote the present values of $H(z)$, $\Omega_{DM}(z)$ and $\Omega_{DE}(z)$ respectively. In particular, we have considered that the universe consists of dark matter and dark energy, and therefore the total density parameter of the universe is $\Omega_{DM0}+\Omega_{DE0}=1.$\\
%%%%%%%%%%%%%%%%%%%%%%%%%%%%%%%%%%%%%%%%%%%%%%%%%%%%%
\par For this choice of $\rho_{DE}(z)$, the {\it deceleration parameter} $q$ evolves as 
\bea\label{eqdp}
q=-\frac{\ddot{a}}{aH^{2}}=-1+\frac{(1+z)}{H(z)}\frac{dH(z)}{dz}~~~~~~~~~~~~~~~~~~~~~~~~~\nonumber \\ 
=-1 + \frac{2\alpha \Omega_{DE0} \frac{z}{(1+z)^2} + 3\Omega_{DM0}(1+z)^3}{2{\Big[\Omega_{DM0}(1+z)^3 + \Omega_{DE0}\Big(1+\alpha {\Big(\frac{z}{1+z}\Big)^2}\Big) \Big]}}
\eea
%%%%%%%%%%%%%%%%%%%%%%%%%%%%%%%%%%%%%%%%%%%%%%%%%%%%%%%%%%%%%%%%%%%%%%%%
For the present model, the dark energy {\it equation of state} (EoS) parameter becomes
\bea
w_{DE}(z)=\frac{p_{DE}}{\rho_{DE}}=\frac{(1+z)\frac{dH^2(z)}{dz} - 3H^2(z)}{3H^2(z)-\rho_{DM}(z)} \nonumber\\
=-1 + \frac{2\alpha z}{3{(1+z)^2} {\Big(1+\alpha {\Big(\frac{z}{1+z}\Big)^2}\Big)}}
\eea
which is independent of the present matter density parameter $\Omega_{DM0}$. Another interesting point regarding the above expression of $w_{DE}(z)$ is that for $\alpha=0$, this behaves exactly like the standard $\Lambda$CDM ($w_{\Lambda}=-1$) model. So, the estimated value of $\alpha$ will indicate whether a cosmological constant or a time evolving dark energy is preferred by cosmological observations.\\
%%%%%%%%%%%%%%%%%%%%%%%%%%%%%%%%%%%%%%%%%%%%%%%%%%%%%%%%%%%%%%%%%%%%%%%%%%%%%%%%%%%%%%%%%%%%%%%%%%
\par In the next section, we shall try to extract the values of the model parameters using the latest cosmological dataset. 
%%%%%%%%%%%%%%%%%%%%%%%%%%%%%%%%%%%%%%%%%%%%%%%%%%%%%
\section{Observational data and fitting method}\label{data}
%%%%%%%%%%%%%%%%%%%%%%%%%%%%%%%%%%%%%%%%%%%%%%%%%
In this section, we have explained the datasets and their analysis method employed to constrain the proposed theoretical model. We have used datasets from the following probes:\\ \\
$\bullet$ {\bf $H(z)$ data:} We have used observational $H(z)$ dataset consisting 41 data points, to probe the nature of dark energy. Among them, 36 data points (10 data points are deduced from the radial BAO size method and 26 data points are obtained from the galaxy differential age method) are compiled by
Meng et al. \cite{hzdataMeng} and 5 new data points of H(z) are obtained from the differential age method by Moresco et al. \cite{hzdataMore}. For the $H(z)$ dataset, the $\chi^2$ is defined as
\be
\chi^2_h (\theta)=\sum^{41}_{i=1} \frac{[h^{obs}(z_i) - h^{th}(z_{i},\theta)]^2}{\sigma^2_{h}(z_i)},~~~h=\frac{H(z)}{H_{0}}
\ee
where, $\theta$ is any model parameter, the superscript ``th" refers to theoretical quantities and superscript ``obs" is for the corresponding observational ones. Also, the uncertainty for normalized $H(z)$ is given by \cite{sigmah,sigmahaam}
\be
\sigma_{h}=\sqrt{\Big(\frac{H^2}{H^4_0}\Big)\sigma^2_{H_{0}} + \frac{\sigma^2_{H}}{H^2_0}}=h \sqrt{\frac{\sigma^2_{H_{0}}}{H^2_0} + \frac{\sigma^2_{H}}{H^2}} 
\ee
where $\sigma_{H_0}$ and $\sigma_{H}$ are the uncertainties in $H_0$ and $H$ respectively. In addition, the present value of $H(z)$ is taken from Ref. \cite{H0}.\\ \\
%%%%%%%%%%%%%%%%%%%%%%%%%%%%%%%%%%%%%%%%%%%%%%%%%%%%%%%%%%%%%%%%%%%%%%%%%%%%%%%%%%%%%
$\bullet$ {\bf SNIa data:} Next, we have incorporated the Union2.1 compilation \cite{u2.1} dataset of total 580 data points with redshift ranging from 0.015 to 1.414. This observations directly measure the {\it distance modulus} of a supernova and its redshift. The relevant $\chi^2$ for the SNIa dataset is defined as \cite{chisn}
\be 
\chi^2_{SN}(\theta)= A(\theta) - \frac{B^2(\theta)}{C(\theta)}
\ee
where $A(\theta)$, $B(\theta)$ and $C(\theta)$ are given by
\bea
A(\theta) = \sum^{580}_{i=1} \frac{[{\mu}^{obs}(z_{i}) - {\mu}^{th}(z_{i},\theta)]^2}{\sigma^2_{\mu}(z_{i})}\\
B(\theta)= \sum^{580}_{i=1} \frac{[{\mu}^{obs}(z_i) - {\mu}^{th}(z_{i},\theta)]}{\sigma^2_{\mu}(z_{i})}
\eea
and
\be 
C(\theta)= \sum^{580}_{i=1} \frac{1}{\sigma^2_{\mu}(z_{i})}
\ee 
%%%%%%%%%%%%%%%%%%%%%%%%%%%%%%%%%%%%%%%%%%%%%%%%%%%%%%%%%%%%%%%%
$\bullet$ {\bf BAO/CMB data:} We have also used BAO  and CMB  measurements data to obtain the BAO/CMB constraints on the model parameters. For BAO data, the results from the 6dFGS Survey measurement at $z=0.106$ \cite{beutler11}, the WiggleZ Dark Energy Survey measurement at $z=0.44,0.6$ and $0.73$ \cite{blake11}, the SDSS DR7 Survey measurement at $z=0.35$ \cite{padmanabhan12} and BOSS CMASS Survey measurement at $z=0.57$ \cite{anderson14} have been used. In addition, we have also used the CMB data derived from the Planck 2015 observations \cite{Planck:2015xua} for the combined analysis TT, TE, EE+lowP+lensing. In this case, the $\chi^2$ function is defined as 
\be
\chi^2_{BAO/CMB}=X^T C^{-1}X 
\ee
where, $X$ and $C^{-1}$ are the transformation matrix and the inverse covariance matrix, respectively \cite{chibc1}. For this dataset, the details of the methodology for obtaining the constraints on model
parameters are described in Refs. \cite{chibc1,chibc2}.\\
%%%%%%%%%%%%%%%%%%%%%%%%%%%%%%%%%%%%%%%%%%%%%%%%%%%%%%%%%%%%%%%%%%%%%
\par One can now use the maximum likelihood method and take the total likelihood function as
\be 
{\cal L}={\rm e}^{-\frac{\chi^2_{i}}{2}}
\ee
where, $\chi^2_{i}=\chi^2_{h}+\chi^2_{SN}+\chi^2_{BAO/CMB}$. The advantage of considering the combined (SNIa+$H(z)$+BAO/CMB) dataset is that they compose an independent dataset, which can help to break the degeneracies between the parameters. Therefore, the combined dataset might also shed light on the cosmological models we aim to investigate. The best-fit corresponds to the model parameters for which the $\chi^2$ (likelihood function) is minimized (maximized). In this analysis, have minimized the $\chi^{2}$ function (say, $\chi^2_{m}$) with respect to the model parameters $\lbrace \Omega_{DM0},\alpha\rbrace$ to obtain their best fit values.  
%%%%%%%%%%%%%%%%%%%%%%%%%%%%%%%%%%%%%%%%%%%%%%%%%%%%%%%%%
\section{Results}\label{result} 
%%%%%%%%%%%%%%%%%%%%%%%%%%%%%%%%%%%%%%%%%%%%%%%%%%%%%%%%%%%%%%%%%%%%%%%%
In this section, we have discussed the results obtained from the $\chi^2$ analysis method using the SNIa, $H(z)$, BAO and CMB datasets. The $1\sigma$ and $2\sigma$ confidence level contours in $\Omega_{DM0}-\alpha$ plane is shown in figure {\ref{figc}. The best-fit values for the model parameters are obtained as $\Omega_{DM0}=0.255$ and $\alpha=-1.883$ (with $\chi^2_{m}=29.04$) for the $H(z)$ dataset. On the other hand, for the SNIa+$H(z)$+BAO/CMB dataset, the corresponding best-fit values are obtained as $\Omega_{DM0}=0.283$ and $\alpha=-1.76$ (with $\chi^2_{m}=610.58$). It should be noted that the best-fit value of $\Omega_{DM0}$ obtained in this work is slightly smaller than the value obtained by the Planck observation \cite{planckr}. We have also found from figure \ref{figc} that the addition of SNIa and BAO/CMB datasets lead to substantially tighter constraints on the model parameters for this model. It has been found that for the best fit model, the dark energy density $\rho_{DE}(z)$, as given in equation (\ref{eans1}), becomes negative at $z\gtrsim 2.8$ and $z\gtrsim 3.1$ for the $H(z)$ and SNIa+$H(z)$+BAO/CMB datasets respectively. The origin of this discrepancy may come from the the choice of $\rho_{DE}(z)$ as well as datasets considered here. However, the situation will be completely different for other choices of $\alpha$. For example, $\rho_{DE}(z)$ becomes finite and positive for $\alpha > -1$. Therefore, this limitation on $\alpha$ is marginally constrained by the datasets analyzed here (see figure {\ref{figc}).\\
%%%%%%%%%%%%%%%%%%%%%%%%%%%%%%%%%%%%%%%%%%%%%%%%
\begin{figure}[ht]
\begin{center}
\includegraphics[width=0.32\textwidth,height=0.18\textheight]{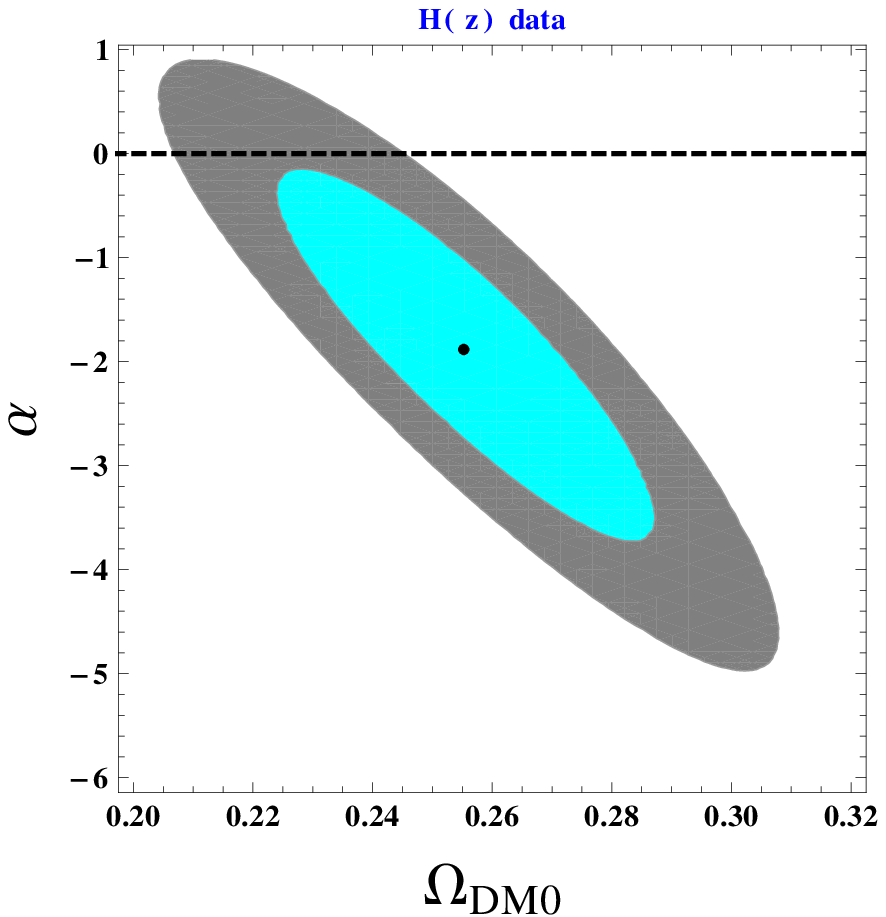}\\
\vspace{7mm}
\includegraphics[width=0.32\textwidth,height=0.18\textheight]{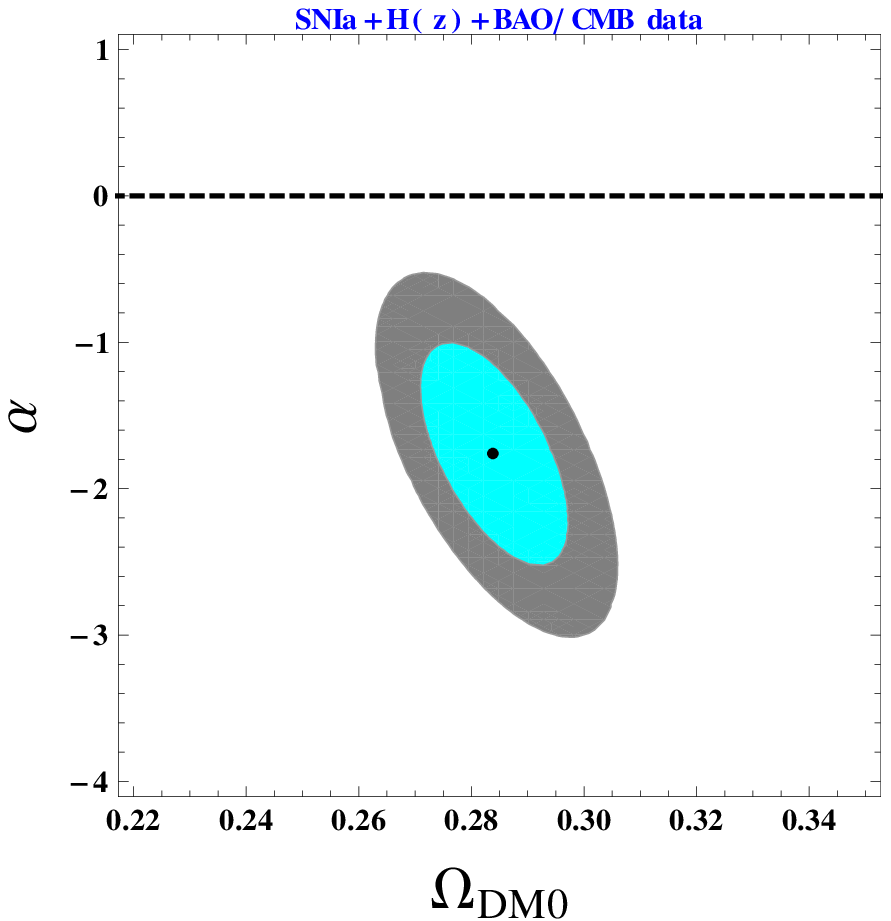}
\caption{Plot of $1\sigma$ (cyan) and $2\sigma$ (gray) confidence contours on $\Omega_{DM0}-\alpha$ parameter space by considering the $H(z)$ (upper panel) and SNIa+$H(z)$+BAO/CMB (lower panel) datasets. In each panel, the black dot represents the best-fit values of ($\Omega_{DM0},\alpha$) and the horizontal dashed line corresponds to the $\Lambda$CDM case ($\rho_{\Lambda}=$constant for $\alpha=0$).}
\label{figc}
\end{center}
\end{figure}
%%%%%%%%%%%%%%%%%%%%%%%%%%%%%%%%%%%%%%%%%%%%%%%%%%%%%%%%%% 
%%%%%%%%%%%%%%%%%%%%%%%%%%%%%%%%%%%%%%%%%%%%%%%%%%%%%%%%
\par As discussed earlier, the model parameters $\alpha$ is a good indicator of deviation of our model from cosmological constant as for $\alpha=0$, the model behaves like the $\Lambda$CDM model. It is observed from figure \ref{figc} that the $\Lambda$CDM model is ruled out at more than $2\sigma$ confidence level by the combined (SNIa+$H(z)$+BAO/CMB) dataset, but it is still in agreement with the  $H(z)$ dataset at the $2\sigma$ confidence level. Figure \ref{figq} shows the evolution of the deceleration parameter $q$ for the best-fit values of $\Omega_{DE0}$ and $\alpha$ arising from the analysis of the $H(z)$ (black curve) and SNIa+$H(z)$+BAO/CMB (red curve) datasets. It is seen from figure \ref{figq} that the universe was decelerated ($q>0$) in the past, began to accelerate at $z\sim 0.84$ (for $H(z)$ data) and $z\sim 0.77$ (for SNIa+$H(z)$+BAO/CMB data), is presently accelerated ($q<0$) but will return to a decelerating phase in the near future. This results are consistent with the results obtained by several authors from different cosmological scenarios \cite{fudp1,fudp2,fudp3,fudp4}.\\
%%%%%%%%%%%%%%%%%%%%%%%%%%%%%%%%%
\par In the upper panel of figure \ref{fighp}, we have shown the best-fit evolution of the dark energy EoS parameter $w_{DE}$ as a function of z for different datasets. It has been found that for each dataset, $w_{DE}(z)$ resembles a $\Lambda$CDM ($w_{\Lambda}=-1$) model at the present epoch (i.e., $z=0$), but, finally, will become positive in the near future. This result is also consistent with the recent works as given in Refs. \cite{arman,liw,magana}, where authors have shown that the cosmic acceleration is currently witnessing its slowing down by using a distinct method. For the sake of completeness, in the lower panel of figure \ref{fighp}, we have plotted the percentage deviation $\bigtriangleup h$ for the above model as compared to a $\Lambda$CDM model, and the corresponding deviation is observed to be $4-5\%$ at low redshifts ($z\sim 0.5$).\\ \\ 
Therefore, the overall dynamic behaviour of our model supports the claims of Valentino et al. \cite{nlcdm1}, Sahni et al. \cite{nlcdm2} and Ding et al. \cite{nlcdm3} that the $\Lambda$CDM model may not be the best description of our universe and also seems to be in agreement with the requirements of String theory \cite{st1,st2,st3}.
%%%%%%%%%%%%%%%%%%%%%%%%%%%%%%%%%%%%%%%%%%%%%%%%
\begin{figure}[ht]
\begin{center}
\includegraphics[width=0.3\textwidth,height=0.18\textheight]{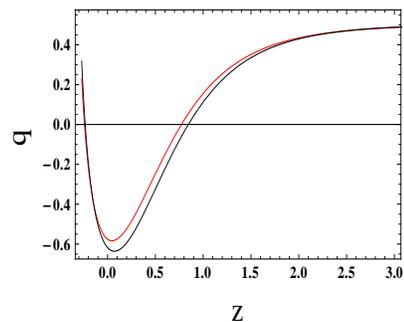}
\caption{The best-fit evolution of the deceleration parameter $q$ is shown for the present model given by equation (\ref{eqdp}). The black curve is for the $H(z)$ dataset while the red one is for the SNIa+$H(z)$+BAO/CMB dataset.}
\label{figq}
\end{center}
\end{figure}
%%%%%%%%%%%%%%%%%%%%%%%%%%%%%%%%%%%%%%%%%%%%%%%%%%%%%%%%%% 
%%%%%%%%%%%%%%%%%%%%%%%%%%%%%%%%%%%%%%%%%%%%%%%%
\begin{figure}[ht]
\begin{center}
\includegraphics[width=0.3\textwidth,height=0.18\textheight]{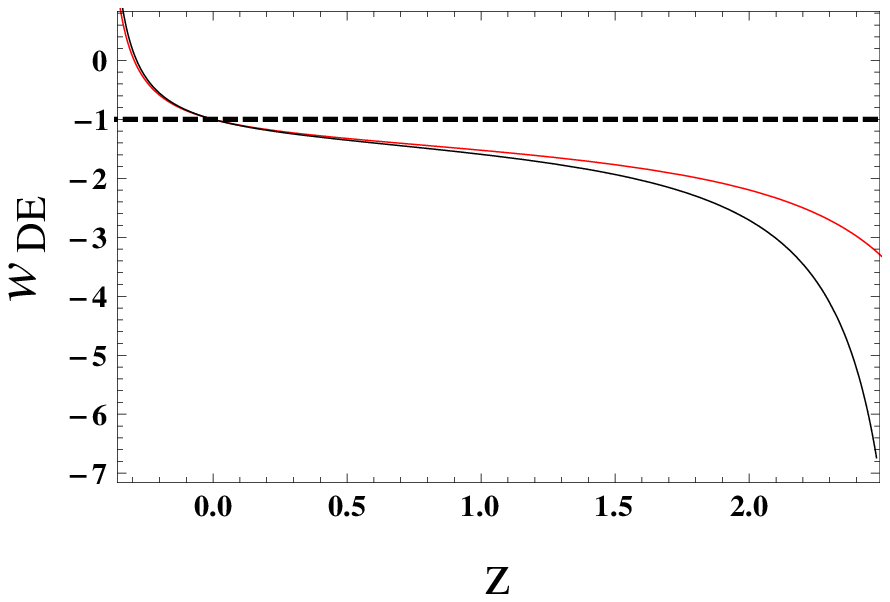}\\
\vspace{7mm}
\includegraphics[width=0.3\textwidth,height=0.18\textheight]{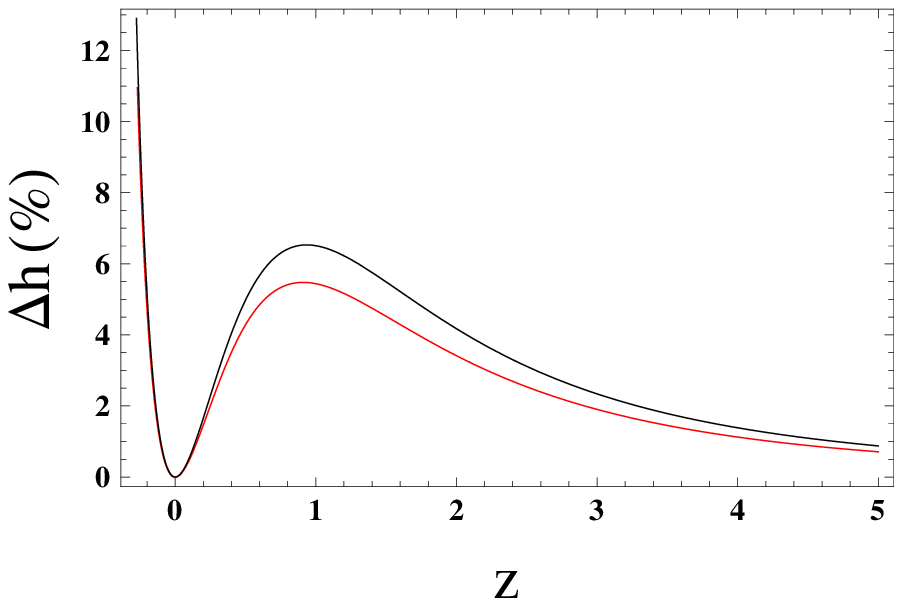}
\caption{Upper panel shows the best-fit evolution of $w_{DE}(z)$ for the present model. The horizontal dashed line indicates the $\Lambda$CDM ($w_{\Lambda}=-1$) case and is shown here for the sake of comparison. Lower panel shows the percentage deviation, $\bigtriangleup h(\%)= \frac{h(z)-h_{\Lambda CDM}(z)}{h_{\Lambda CDM}(z)}\times 100$. In each panel, the black and red curves represent the result from the $H(z)$ and SNIa+$H(z)$+BAO/CMB datasets respectively. }
\label{fighp}
\end{center}
\end{figure}
%%%%%%%%%%%%%%%%%%%%%%%%%%%%%%%%%%%%%%%%%%%%%%%%%%%%%%%%%%
%%%%%%%%%%%%%%%%%%%%%%%%%%%%%%%%%%%%%%%%%%%%%%%%%%%%%%%%%%%%%%
\section{Conclusions}\label{conclusion}
%%%%%%%%%%%%%%%%%%%%%%%%%%%%%%%%%%%%%%%%%%%%%%%%%%%%%%%
In this work, we have studied the dynamics of accelerating scenario of the universe by considering one specific parameterization of the dark energy density $\rho_{DE}$ and from this we have obtained analytical solutions for different cosmological parameters. As we have mentioned before, the new parametrization of $\rho_{DE}$, given by equation (\ref{eans1}), reduces to other popular dark energy models for different choices of $\alpha$ and $n$. Of course, $n$ may assume any real value, but we have found that for $n=2$, the expression of the dark energy density proposed in equation (\ref{eans1}) gives interesting consequences. So, in the present work, we have confined our investigation to $n=2$ only. We have used the recent Hubble parameter dataset along with the estimation of the local Hubble parameter value as well as the standard dark energy probes, such as the SNIa, BAO and CMB measurements to constrain different parameters of our model. It has been found that we need $\alpha>-1$ to ensure the finite and positive value of $\rho_{DE}$, and the model seems to be marginally consistent with the observational datasets analyzed here. \\
%%%%%%%%%%%%%%%%%%%%%%%%%%%%%%%%%%%%%%%%%%%%%%%%%%%%
\par In summary, our analysis predicts a transient accelerating phase, in which the universe was decelerated ($q>0$) in the past, began to accelerate at redshift $z<1$, is presently accelerated ($q<0$), but will return to a decelerating phase in the near future. This overall dynamic behaviour is much different from the standard $\Lambda$CDM scenario. Hence, the present model supports the claims of several authors \cite{nlcdm1,nlcdm2,nlcdm3} that the $\Lambda$CDM model may not be the best description of our universe. Therefore, this specific dark energy model, with a transient accelerating phase and $\alpha \neq 0$, can be considered as an alternative for the $\Lambda$CDM model. 
%%%%%%%%%%%%%%%%%%%%%%%%%%%%%%%%%%%%%%%%%%%%%%%%%%%%%%%%%%%%%%%%%%%%%
%%%%%%%%%%%%%%%%%%%%%%%%%%%%%%%%%%%%%%%%%%%%%%%%
\section{Acknowledgments}
The author would like to thank the two anonymous referees for useful comments and suggestions. The author acknowledges the financial support from the Science and Engineering Research Board (SERB), Government of India through National Post-Doctoral Fellowship Scheme (File No: PDF/2017/000308). The author also wishes to thank the Inter University Center for Astronomy and Astrophysics (IUCAA), Pune for their warm hospitality as a part of the work was done during a visit. 
%%%%%%%%%%%%%%%%%%%%%%%%%%%%%%%%%%%%%%%%%%%%%%%%%%%%%%%%%%%%%%%%%%%%%
%%%%%%%%%%%%%%%%%%%%%%%%%%%%%%%%%%%%%%%%%%%%%%%%%%%%%%%%%%%%%%%

\end{document}